\documentclass{appolb}
\usepackage{float}
\usepackage[caption = false]{subfig}
\usepackage[final]{graphicx}
\usepackage{cancel}
\usepackage{amsmath}
\usepackage{hyperref}
\hypersetup{colorlinks=false, linkcolor=black, filecolor=black, urlcolor=black}

\usepackage{tikz}
\usepackage{amsmath,amsfonts,amssymb,amsthm}
\allowdisplaybreaks
\newcommand{\bea}{\begin{eqnarray}} % \newcommand{\bea} {\begin{eqnarray*}}
\newcommand{\eea}{ \end{eqnarray}} % \newcommand{\end{eqnarray}} {\end{eqnarray*}  }
\newcommand{\ba}{\begin{eqnarray*}}
\newcommand{\ea}{  \end{eqnarray*}}

\newcommand{\beq}{\begin{equation*} }
\newcommand{\eeq}{\end{equation*}  }

\newcommand{\bqa}{\begin{eqnarray*} }
\newcommand{\eqa}{\end{eqnarray*}}

\def\beq{\begin{equation}}
\def\eeq{\end{equation}}
\def\beqa{\begin{eqnarray}}
\def\eeqa{\end{eqnarray}}
\def\ben{\begin{enumerate}}
\def\een{\end{enumerate}}
\def\bit{\begin{itemize}}
\def\eit{\end{itemize}}

\def\mcC{\mathcal{C}}
\def\mcJ{\mathcal{J}}
\def\mcA{\mathcal{A}}
\DeclareMathOperator*{\res}{Res}

\def\MB{{\textsf{MB\/}}}
\def\MBresolve{{\textsf{MBresolve.m\/}}}

\def \ar      {\texttt{AMBRE}{}}

\def \mbn      {\texttt{MBnumerics.m}{}}

\def \mb {\texttt{MB}{}}

\def \mbm  {\texttt{MB.m}{}}

\begin{document}
% \eqsec  % uncomment this line to get equations numbered by (sec.num)
\title{ %Mellin and Barnes meet in Minkowskian kinematics. \\
\vspace*{-10mm}
{\hfill  \tt KW 17-001}
\vspace*{10mm}
\\
New prospects for the numerical calculation \\of Mellin-Barnes integrals in Minkowskian kinematics
%
% you can use '\\' to break lines
}

\author{Ievgen Dubovyk
\address{
II. Institut f{\"u}r Theoretische Physik, Universit{\"a}t Hamburg,\\
Luruper Chaussee 149, 
22761 Hamburg,  Germany} 
%\\ \vspace{0.3cm}
%{Ayres~Freitas}
%\address{Pittsburgh Particle physics, Astrophysics \& Cosmology Center
%(PITT PACC),\\ Department of Physics \& Astronomy, University of Pittsburgh, PA 15260, USA}
\\ \vspace{0.3cm}
{Janusz Gluza{\thanks{Presented by J.Gluza at the Epiphany Cracow conference 2017}}, Tomasz Jeli\'nski, Tord Riemann}
 \address{Institute of Physics, University of Silesia, \\
Uniwersytecka 4, 40-007 Katowice, Poland}
%{Tord Riemann}
%\address{Institute of Physics, University of Silesia, 
%Uniwersytecka 4, 40-007 Katowice, Poland}
\\ \vspace{0.3cm}
{Johann Usovitsch}
\address{Institut f{\"u}r Physik, Humboldt-Universit{\"a}t zu Berlin, 12489 Berlin, Germany}
}

%\author{Krzysztof Bielas\thanks{Presented by K. Bielas at the International Conference of Theoretical Physics ``Matter To The Deepest'', Ustron 2013}, Ievgen Dubovyk, Janusz Gluza}
%\address{Institute of Physics, University of Silesia, Uniwersytecka 4, PL-40-007 Katowice, Poland}\\
%\author{Tord Riemann}
%\address{Deutsches Elektronen-Synchrotron, DESY, Platanenallee 6, 15738 Zeuthen, Germany}

\maketitle
\begin{abstract}
 During the last several years remarkable progress has been made in numerical calculations of dimensionally regulated multi-loop 
 Feynman diagrams using Mellin-Barnes (MB) representations.
 The bottlenecks were non-planar diagrams and Minkowskian kinematics.
 The method has been proved to work in highly non-trivial physical application (two-loop electroweak bosonic corrections to the $Z \to b \bar{{b}}$ decay), and cross-checked with the sector decomposition (SD) approach.  In fact, both approaches have their pros and cons. In calculation of multidimensional integrals, depending on masses and scales involved, they are complementary. 
 A powerful top-bottom approach to the numerical integration of multidimensional \MB{} integrals is automatized in the \MB{} suite 
 AMBRE/MB/ MBtools/MBnumerics/CUBA.
 Key elements are a dedicated use of the Cheng-Wu theorem for  non-planar topologies and of shifts and deformations of the integration 
 contours.  
 An alternative bottom-up approach starting with complex 1-dimensional \MB-integrals, based 
 on the exploration of steepest descent integration contours in Minkowskian kinematics, is also discussed.
 Short and long term prospects of the \MB{}-method for multi-loop applications to LHC- and LC-physics are discussed.   
%Two approaches are discussed. 
\end{abstract}
\PACS{02.70.Wz, 12.15.Lk,12.38.Bx}
  
\section{Introduction}

Historically a concept of Feynman diagrams was presented for the first time at a special by-invitation-only meeting at the Pocono Manor Inn 
in Pennsylvania in 1948 by Feynman as an alternative to procedures of perturbative calculations in QED 
\cite{Schweber:1994qa-mod,Kaiser:941915}. The idea was systematically treated for the first time by Dyson in his two seminal papers 
\cite{Dyson:1949bp,Dyson:1949ha} followed by Feynman himself \cite{Feynman:1949hz,Feynman:1949zx}\footnote{That is why initially it was 
being called the Feynman-Dyson approach to QED.}. Integrals which stand behind the diagrams are, together with a renormalization procedure, 
in the core of the technical difficulties, 
which increase with the number of "legs" 
and "loops" involved in calculation of contemporary QCD and electroweak processes. It is clear that steady progress in particle physics 
needs new ideas and crafting ever-changing theoretical tools and techniques of calculations.
  
  In the following the \MB{}-suite will be described to some detail.
  It comprises several tools for dimensionally regulated Feynman integrals in the momentum space:
 (i) Transform them into Feynman integrals expressed by
  Feynman parameters (textbook knowledge);
  (ii) Use AMBRE
  \cite{Gluza:2007rt,Gluza:2010rn,Dubovyk:2016ocz,Katowice-CAS:2007-url} ---
  transform them into Mellin-Barns integrals, valid at initial parameters
  which include a finite shift $\epsilon$ of dimension, $d=4-2\epsilon$, and 
  with original
  integration paths parallel to the imaginary axis;
  (iii) Use \mbm{} or \MBresolve{} \cite{Czakon:2005rk,Smirnov:2009up} ---
  perform an analytical continuation in $\epsilon$,
  approaching small $\epsilon$ and
  (iv) --- expand the Mellin-Barnes integrals as series in small $\epsilon$; 
  (v) Use barnesroutines  \cite{mbtools-url}
   --- perform simplifications using Barnes lemmas.
  (vi)
  At this stage the original representation of the Feynman integral in terms of several finite MB-integrals has been formulated.
  One may now start to calculate them, either analytically or numerically, or in a mixed approach.
  In sufficiently complicated situations, only numerics can be applied.
  (vii) 
  Use  \mbn{} \cite{mbnum}
%,Usovitsch:201606} 
to perform  parametric integrations of the MB-integrals along the paths defined 
  in step (iii), thereby applying a variety of techniques: integration variable transformations, reparameterizations, contour 
  deformations, contour shifts and whatsoever.
%  In not too complicated cases, MBnum of  MB \cite{MB,Czakon:2005rk} gives already quote good results.
  For the parameter integrations, CUHRE of the package CUBA \cite{Hahn:2004fe} is used.
 % {\bf Johann, what else is used?}
  To some extent, we gave some descriptions of details before \cite{Dubovyk:2016ocz,Dubovyk:2016zok,Dubovyk:2016aqv}.
  
 In this article we focus on purely numerical approaches to Feynman integrals developed in last few years beyond one-loop (NLO) perturbation.
 One is faced with several technical obstacles.
 There are infrared singularities. We know about two methods to treat them properly without limitations. One is the MB-method, the other one 
 sector decomposition \cite{Smirnov:2013eza,Borowka:2015mxa}, which is also numerical. 
 We aim at direct calculations in Minkowskian kinematics, which presents serious convergence problems but is crucial for production 
 processes at high energy accelerators like LHC and LC. 
 No doubt that framed with powerful fast, stable, accurate and universal 
 software, direct numerical calculations will become necessary for practical applications on mass scale, similarly as it happened at the NLO  level\footnote{To solve the integrals, 
 analytical methods can be used, though they exhibit natural limitations when sophisticated integrals with many parameters appear. Such a situation takes place in gauge theories, like in the  electroweak-QCD Standard Model.
 However, concerning analytical approaches to Feynman integrals,  we should especially appreciate recent progress in differential equation method \cite{Kotikov:1990kg,Kotikov:1991hm,Kotikov:1991pm,Remiddi:1997ny}, which got a push in 2013 \cite{Henn:2013pwa} followed by latest corresponding software and ideas \cite{Ablinger:2015tua,Prausa:2017ltv,Gituliar:2017vzm,vonManteuffel:2017hms,Adams:2017tga}. Here further progress in developing integration-by-parts (IBP) concepts is also very important \cite{Georgoudis:2016wff}.
  }, see {\rm
   FeynArts/FormCalc}~\cite{Hahn:2000kx,Nejad:2013ina},
 {\rm  CutTools}~\cite{Ossola:2007ax}, {\rm
   Blackhat}~\cite{Berger:2008sj}, {\rm
   Helac-1loop}~\cite{vanHameren:2009dr}, {\rm
   NGluon}~\cite{Badger:2010nx}, {\rm Samurai}~\cite{Mastrolia:2010nb},
 {\rm Madloop}~\cite{Hirschi:2011pa}, {\rm GoSam}~\cite{Cullen:2011ac},
 {\rm PJFry}~\cite{Fleischer:2012et}, {\rm OpenLoops}~\cite{Cascioli:2011va} and  \cite{vanOldenborgh:1990yc,vanHameren:2010cp,Ellis:2007qk,Denner:2016kdg}.

In this article we discuss the state of the art of purely numerical approaches to multiloop integrals, focusing on the Mellin-Barnes 
method.  We show the first completed and non-trivial application in cutting-edge physical calculations using the \MB{}-suite followed by 
further perspectives. 
 
\section{Numerical concepts beyond NLO level}

Fully numerical techniques for the evaluation of two- and higher-loop integrals need
the  extraction of ultraviolet, infrared and
collinear singularities. On top of that, they must be numerically stable and efficient.  
A qualitative comparison of different numerical integration techniques for 
Feynman parameter integration of massive multi-loop integrals
can be found in \cite{Smirnov:2002,Steinhauser:2002rq,Anastasiou:2005cb,Freitas:2016sty}. The main methods are dispersion relations, 
Bernstein-Tkachov method, differential equations, subtraction terms, the SD and MB methods. Here we will discuss the last two, specifically 
focusing on MB. There are presently only few public programs for the numerical integration of integrals beyond NLO level. {\rm 
NICODEMOS}~\cite{Freitas:2012iu} is based on contour deformations.
There are also complete programs 
dedicated specifically to the precise calculation of two-loop self-energy diagrams \cite{Martin:2005qm,Caffo:2008aw}.
However, the most advanced and universal programs are based on the SD or MB approaches:
Sector decomposition, developed into  independent packages (present versions) {\rm Fiesta 4}~\cite{Smirnov:2015mct} and {\rm SecDec 
3}~\cite{Borowka:2015mxa} followed by pySecDec \cite{Borowka:2017idc};
\mbm{} \cite{Czakon:2005rk} and \MBresolve{} \cite{Smirnov:2009up} packages extract $\epsilon$-singularities in dimensional regularization 
of MB multiloop integrals, offering also possibilities of numerical integrations in Euclidean kinematics, which is relatively simple as no 
physical branch cuts are present there. It was used intensively in the past to cross check numerically analytical results for multiloop 
integrals. In next section we will discuss new ideas for making possible \MB numerical integrations directly in the physical region. 

%\cite{sec} and Mellin-Barnes representations \cite{mb,mb2,mellinbarnes}. These methods proved to be also useful in direct numerical calculations.

% Two powerful approaches for a straightforward extraction of $1/(4-D)$ poles in dimensional regularization are

\section{Numerical integrations of \MB{} integrals in Minkowskian region}

%MB.m{\tiny{{[M. Czakon]}}} (cross-checks for analytical calculation of MIs in Euclidean region). Used in many projects.
% {First application: Bhabha massive QED 2-loop:} 

%\subsection{Mellin-Barnes Feynman integrals}

The Mellin-Barnes transformation of Feynman multidimensional integrals to multivariable complex plane 
integrations \cite{mellin1895,barnes1900} has been used in many particle physics calculations.
% by basic relation in which sum of each two terms in some expression can be changed into integration over complex variable. 
In the first applications \cite{Bergere:1973fq,uss1975} this kind of transformation has been applied  directly to propagators in the loop 
integrals, changing $"\rm {momenta}^2-\rm{mass}^2"$ terms into ratios of momenta and masses in complex plane.
Nowadays, a more efficient and systematic treatment of multiloop integrals goes  by expressing Feynman integrals by the Symanzik polynomials 
$F$ and $U$ \cite{Smirnov:1999gc,Tausk:1999vh,Smirnov:2004}, for which the general \MB{} formula is applied
\begin{eqnarray}
  \frac{1}{(A_{1}+ \ldots +A_{n})^{\lambda}}&=&
  \frac{1}{\Gamma (\lambda)}\frac{1}{(2 \pi i)^{n-1}}
  \int_{c-i \infty}^{c+i \infty} \dots \int_{c-i \infty}^{c+i \infty}
  dz_{2} \ldots dz_{n}
  \prod_{i=2}^{n} A_{i}^{z_{i}}
%===================
  \nonumber\\
  &\times & A_{1}^{-\lambda -z_{2}- \ldots -z_{n}}
  \Gamma (\lambda +z_{2}+ \ldots +z_{n})
  \prod_{i=2}^{n} \Gamma (-z_{i}).
\label{mbeq}
\end{eqnarray}

As we can see, $n$ additive terms lead to $n-1$ complex integrals. The $A_i$ terms correspond to kinematical parameters of the integral.
A typical simple example is the 1-dimensional singular part  of the 1-loop massive QED vertex  
\cite{Czakon:2005rk,Dubovyk:2016ocz,Gluza:2016fwh} $\sim \int dz
%\limits_{-\frac{1}{2}-i \infty}^{-\frac{1}{2}+i \infty} 
%\frac{dz}{2\pi i}~~
%\underbrace{
(-s)^{-z}
%}_{\bf {Part\; I}}
%\overbrace{
%\frac{
\Gamma^3(-z)\Gamma(1+z) \Gamma^{-1}(-2z). 
%}}^{\bf {Part\; II}}
$
Choosing properly the contour of integration can make the annoying oscillatory behavior of the term $(-s)^{-z}$ small and controllable (for 
$s>0$, so Minkowskian kinematic). Furthermore, Gamma functions $\Gamma$ exhibit singularities either, and make the task of integral 
evaluations highly non-trivial.  

The construction of \MB{} integrals through Symanzik polynomials is automatized in the \ar{} project 
\cite{Gluza:2007rt,Gluza:2010rn,Dubovyk:2016ocz,Katowice-CAS:2007-url}.
Using it with \mbm{} or \MBresolve{}, IR and UV divergencies can be extracted and regulated multidimensional \MB integrals are obtained \cite{Smirnov:2004ym}. 
On the webpage \cite{mbtools-url} more auxiliary packages with examples related to \MB{} calculations can be found.

The first serious trial directed to the numerical integration of \mb{} integrals in Minkowskian space-time was undertaken in 
\cite{Freitas:2010nx}. 
The method developed there is based on simultaneous rotations of integration paths for all variables by the same angle in the complex plane
  and has been applied successfully to the 
calculation of
two-loop diagrams with triangle fermion subloops for the $Z \to b\bar{b}$  formfactor  \cite{Freitas:2012sy}. 
Another interesting numerical application 
 of \mb{} integrals for phase space integrations can be found in \cite{Somogyi:2011ir} and \cite{Anastasiou:2007qb,Anastasiou:2013srw}. 
There some parametric integrals are considered and transformations of \mb{} integrals into Dirac delta constraints have been explored.
\\ 
Now we will present recent developments. First we describe a Top-Bottom approach in which the \mbn{} package  deals with multidimensional 
\MB{} integrals; it was described partly in \cite{Dubovyk:2016ocz} and applied in \cite{Dubovyk:2016aqv}. Another Bottom-Top approach is at 
the exploratory stage; optimal complex contours of \MB{} integrations are worked out systematically for one-dimensional \MB{} 
integrals \cite{Gluza:2016fwh}.

\subsection{Top-Bottom approach - shifts, deformations and \mbn}

\begin{figure}[h!]
\begin{center}
\hspace*{-.5cm}
\includegraphics[width=.7\textwidth]{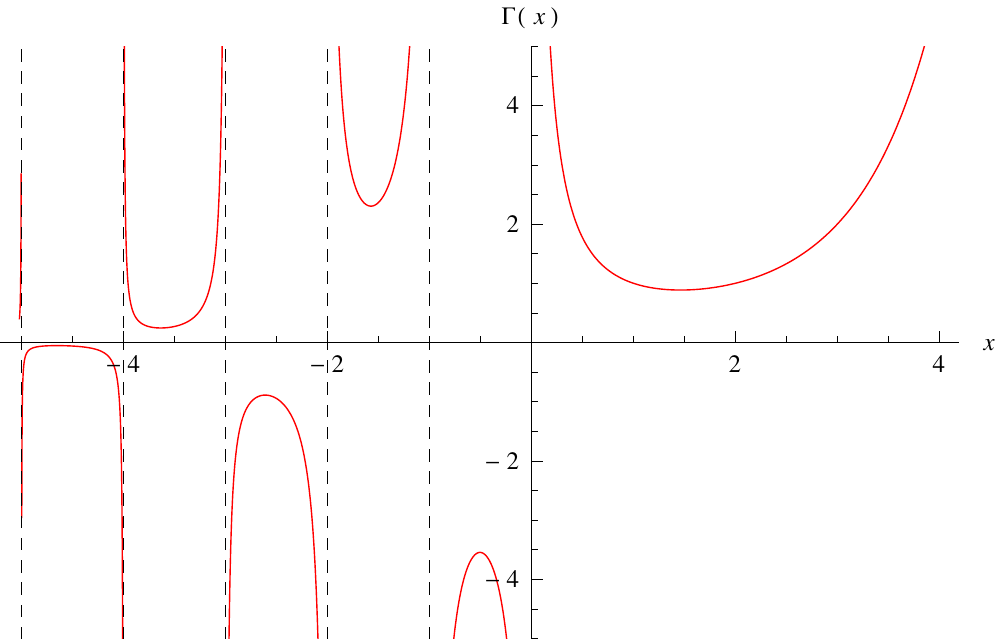}
\end{center}
  \caption{Gamma function defined as 
  $\Gamma (z)=\int_{0}^{\infty} t^{z-1} e^{-t}dt$, $x=\Re[z]$. For more details, see \cite{Barnes:zbMATH02640947,Whittaker:1965}.}
  \label{figregamma}
\end{figure}

As we can see from (\ref{mbeq}), Gamma functions are there \cite{barnes1900}. In Fig.~\ref{figregamma} the real part of the $\Gamma[z]$ 
function is sketched. It is regular in positive $\Re[z]$ and has singularities for integer negative $\Re[z]$. 

\begin{figure}[h!]
\centering
\subfloat[]{\includegraphics[width=.9\textwidth]{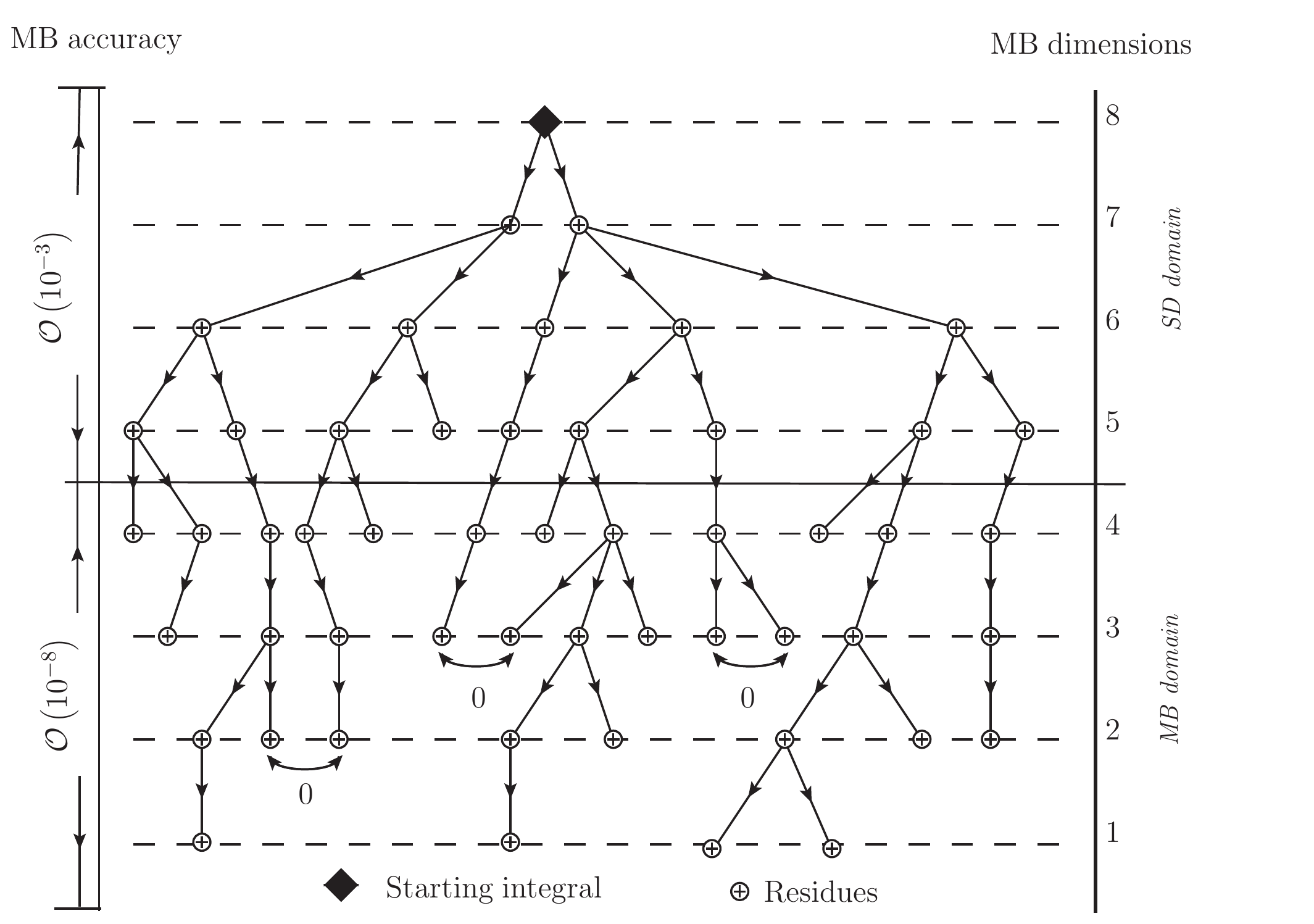}} 
%\subfloat[]{\includegraphics[width=0.35\textwidth]{figs/soft_v}}\    
%\subfloat[]{\includegraphics[width=0.35\textwidth]{figs/0hxwxt0z_b}}
%\subfloat[]{\includegraphics[width=0.35\textwidth]{figs/0hxwxtxz_v}} 
\caption{A possible scenario for the calculation of some 8-dimensional \MB{} integral. Lower dimensionality \MB{} integrals result from 
shifting   complex variables of the integral by integers, as explained in the text.
In  \cite{Dubovyk:2016aqv} all integrals of dimension less than 5 were calculated this way with \mbn{} \cite{mbnum} and high accuracy, remaining integrals were treated with the same accuracy by the SD method. However, as a basic cross-check, less digits could be obtained for all integrals using both methods.}
\label{scheme}
\end{figure}

Note also that at the negative axis between the pole positions, the integrand  becomes smaller in its absolute value for 
the function evaluated at an argument further away from the origin. In addition, for a pole crossed by an argument shift,
one has to add the corresponding residue which by itself is also an 
integral, but will have a dimension less than the original one.
Repeating the procedure for several integration variables, 
the original MB integral gets replaced by several lower-dimensional integrals, and the original one with a shifted integration path.
In the end, the (module of the) resulting contribution of the original integral after shifts can be made smaller than the desired accuracy 
of the calculation. 
In effect, the procedure constructs a sum over a finite number of residues with a controlled remainder. This procedure of 
shifts is implemented in \mbn{} \cite{mbnum}. Some other important features of the 
procedure like contour deformations and mappings of integrated variables into finite intervals have been discussed 
in \cite{Dubovyk:2016ocz,Dubovyk:2016zok}. Fig.\ref{scheme} sketches roughly the idea.

\begin{figure}[h!]
\centering
\subfloat[]{\includegraphics[width=0.35\textwidth]{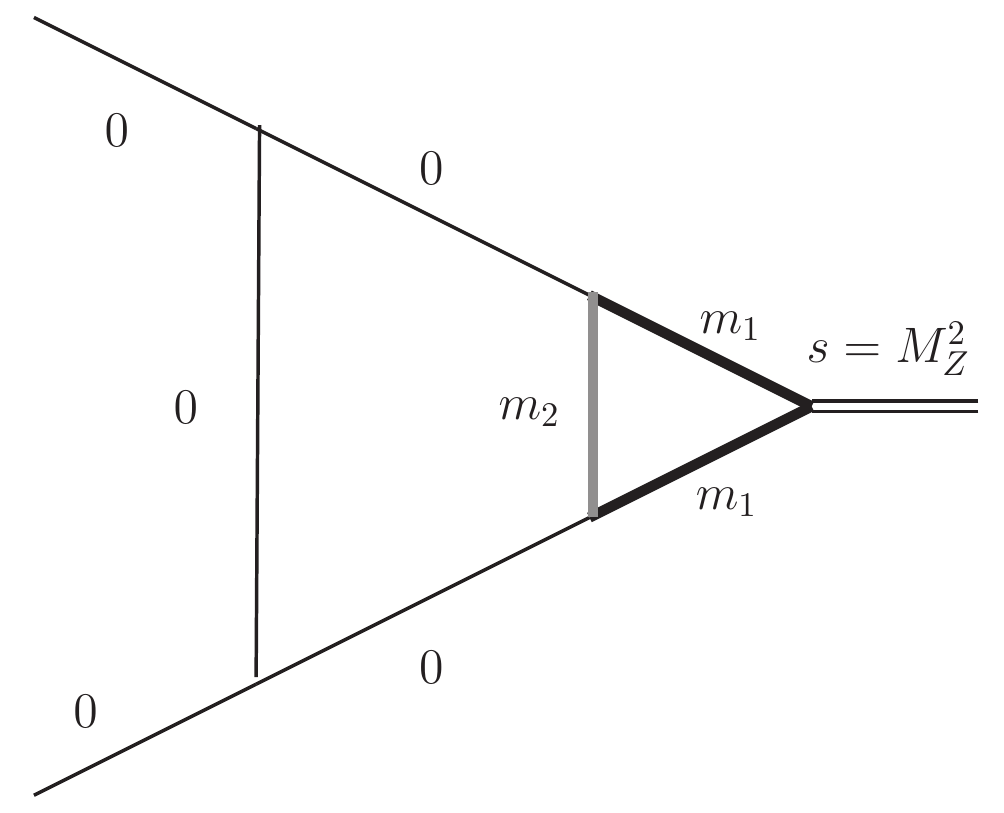}} 
\subfloat[]{\includegraphics[width=0.35\textwidth]{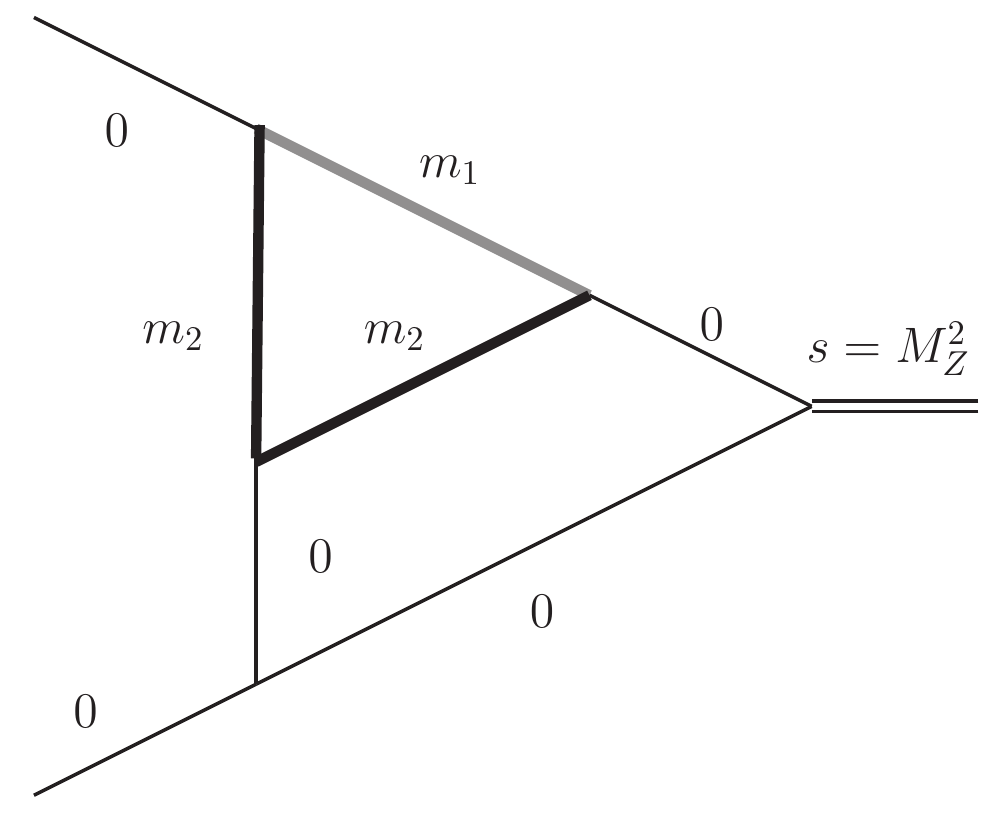}}\    
\subfloat[]{\includegraphics[width=0.35\textwidth]{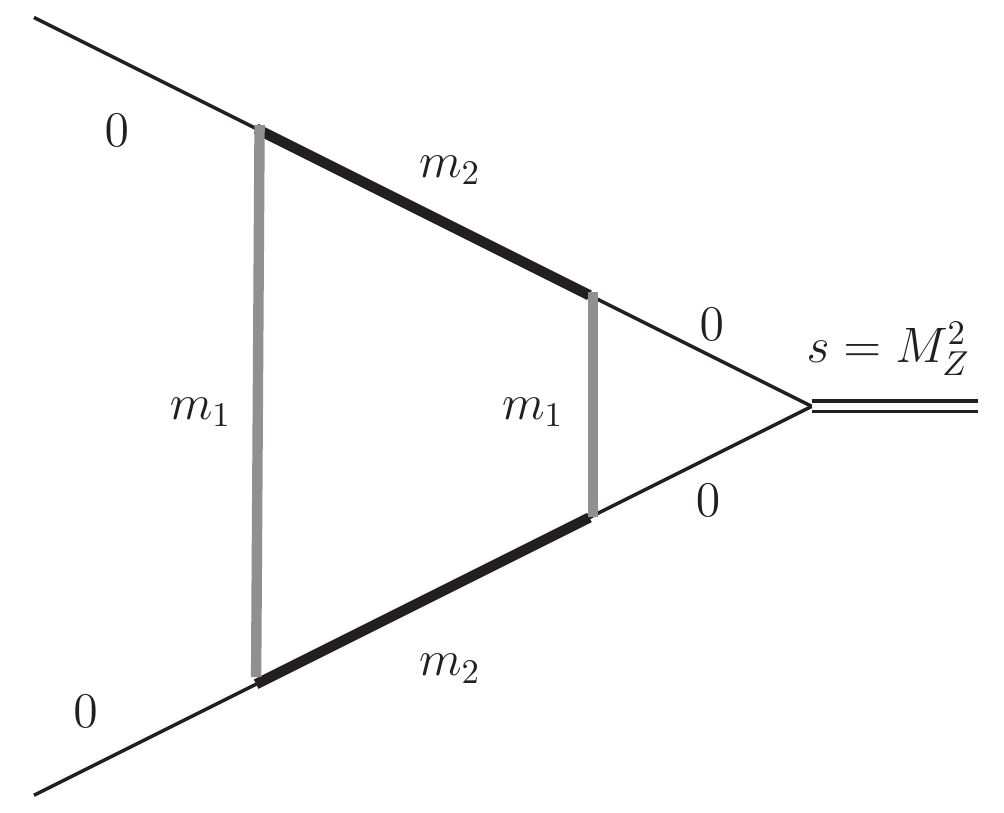}}
\subfloat[]{\includegraphics[width=0.35\textwidth]{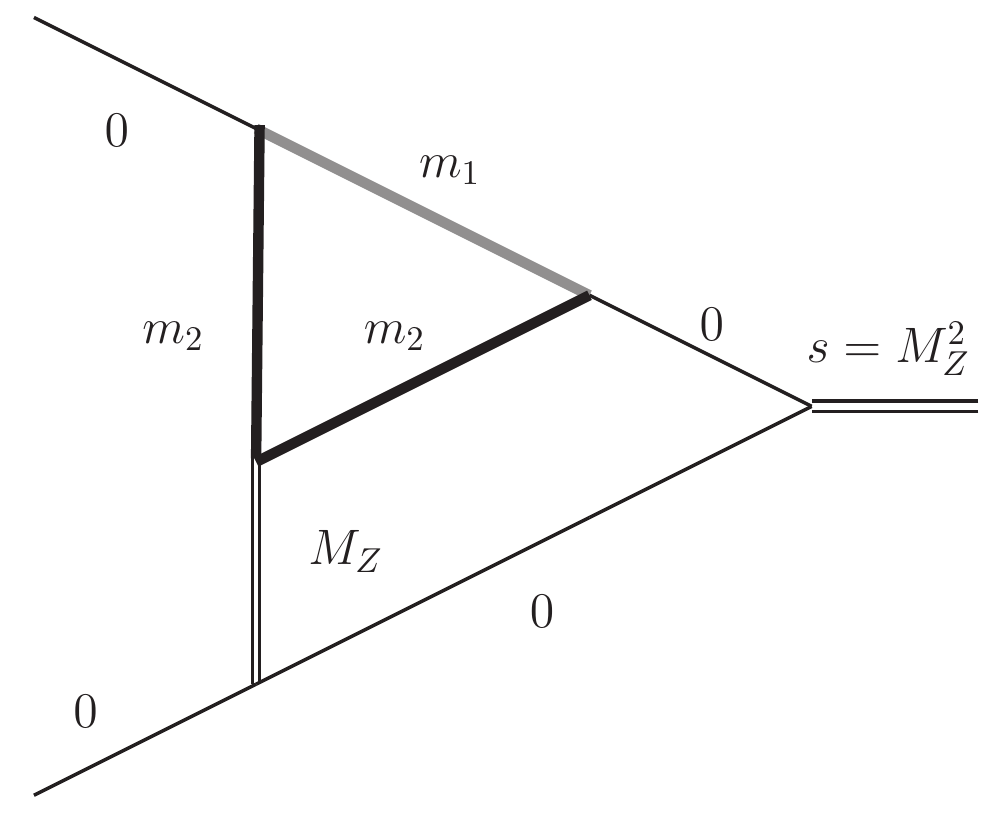}} 
\caption{An example of diagrams appearing in the calculation of $\sin^2\theta_{\rm eff}^{\rm b}$ \cite{Dubovyk:2016aqv} for which \MB{} and 
{\rm SD} methods have been applied.
The \mb{} representations have up to 4-dimensional integrals, to be taken at the $Z$ boson mass shell, $s=M_Z^2$. A numerical accuracy at 
the level $\mathcal{O}(10^{-8})$ was 
achieved for them only  with the \mb{} approach. For a general $s$-dependence the situation is the same.}
\label{difcltcases}
\end{figure}

\captionsetup[subfigure]{labelformat=empty}
In the project \cite{Dubovyk:2016aqv} we derived Mellin-Barnes representations for all integrals, which had up to eight dimensions. 
For a cross-check, each integral was computed with \MB{} and SD techniques. There are only few classes of diagrams for which eight digits 
could not be achieved with both methods, an example is given in Fig.\ref{difcltcases}; for further discussion, see 
\cite{Dubovyk:2016ocz,Dubovyk:2016zok}. 
These diagrams have high order divergiences and an application of the
sector decomposition approach leads to numerical problems related to an
accuracy and time consumption. In contrast to this, the corresponding \MB{}
integrals can be computed with reasonable computer time resources.

Typically, for integrals which involve many masses, SD fits better while the \MB{} method works out perfectly for more "massless" diagrams. 
Thus, the \MB{} suite and the sector decomposition techniques are to a large extent complementary \cite{Gluza:2010rn} and both numerical 
methods can be successfully  explored together in cutting-edge physical problems.

%For instance, at two loops level, in the $Z b b$ case, 4-dim and lower \mb{} integrals appear for diagrams with very few internal 
%masses and external virtualities, see Fig.\ref{difcltcases}.

\captionsetup[subfigure]{labelformat=empty}
In \cite{Dubovyk:2016aqv} for \MB{} and planar diagrams  the newest version \ar{} v2.1 \cite{Katowice-CAS:2007-url} is used, for non-planar 
diagrams it is \ar{} v3.1 \cite{Katowice-CAS:2007-url}. Planarity of diagrams is controled automatically with the {\rm PlanarityTest.m} 
package \cite{Bielas:2013v11,Bielas:2013rja}. Numerical results have been obtained using \mbn{} \cite{mbnum}. As it is demonstrated in 
Fig.~\ref{scheme}, the shifts accumulate at each new iteration many residues, until the desired accuracy is reached. It is worth noting that 
the integration error of \mbn{} is mostly dominated by the collection of residues which have fast convergence. For higher-dimensional 
integrals,  \mbn{} collects more residues. The resulting error from all residues is determined by Pythagorean addition. 
In Fig.~\ref{scheme} the double arrows with zero marks denote pairs of residues which are identified to finally cancel exactly. To identify  
such pairs to a high accuracy, \mbn{} performs the integration of the corresponding candidates at a different kinematical point where a  
high numerical accuracy is reached. If then the integrals agree up to a sign, \mbn{} sums them up to zero. This is only one example of 
many numerical problems which have been solved in the \mbn{} algorithm, in order to get highly accurate numerical results in the 
Minkowskian region. The package is yet under development, and our present estimation is that in the near future even 12-dimensional \MB{} 
integrals can be touched -- e.g. pentaboxes.

%The present situation is that ... 
%{\bf EVGEN: more on the most difficult 1dim, 2dim, 3dim cases figs
%Noting complementarity of secdec and mb approaches.}

%The present situation is that the application of this method to physical kinematics is limited by the dimensionality of the \mb{} integrals, nonetheless, interesting physical processes can be investigated with the method.
%But this doesn't restrict an area of its applications to physical problems. 

%\clearpage

%\begin{eqnarray*}
%&&\int dz_1 \ldots dz_k \ldots I(\ldots, {Re[z_k]+n}+Im[z_k],\ldots) \hspace*{2cm} I_{orig}\\ &=&  Residue[  \int dz_1 \ldots \cancel{dz_{k}} \ldots  I]_{Re[z_k]+n}  \hspace*{4cm} I_{Res} \\
%&+& \int dz_1 \ldots dz_k \ldots I(\ldots, {Re[z_k]+(n+1)}+Im[z_k],\ldots)  \hspace*{1cm} I_{new}
%\end{eqnarray*}

%Residues {\bf lower} dimensionality of original MB integrals.
%Integral after passing a pole (proper shifts) {\bf can be made smaller}.

% \begin{figure}[h!]
%  \centering
%   \vspace{-4.5cm}
%    \includegraphics[width=\textwidth]{scheme1}
%    \label{scheme}
%    \caption{JOHANN: Explanation/description and better sketch}
%  \end{figure}

\subsection{Bottom-Top {\tt MBDE} approach - optimal steepest descent contours}
In a nutshell, this  is a { stationary phase method} leading to optimal {steepest descent integration contours}. They can be
found using {Lefschetz thimbles} (exact contours) or their {Pad\'e} approximation \cite{Gluza:2016fwh}.
 
Lefschetz thimbles (LT) are a fascinating subject, crossing many issues
like behaviour of LT in presence of poles, singularities and branch cuts, behaviour in the complex infinity, Stokes phenomenon, relation to 
relative homology of a punctured Riemann sphere, etc. It can be applied e.g. to the 
analytical continuation of 3d Chern-Simons theory, QCD with chemical potential, resurgence theory, counting master integrals or repulsive Hubbard model. 
Still, applying this method to the numerical evaluation of \MB{} integrals is at the exploratory stage and an effective and general 
determination of multivariate \MB{} contours must be worked out yet in more detail.

In this section we present the main idea as 
%a novel 
an alternative 
approach to the numerical computation of \MB{} integrals, starting from the bottom, the lowest one-dimensional \MB{} integrals,  
in both Euclidean ($s<0$) and Minkowski $(s>0)$ regions. These cases have been explored in fine details in \cite{Gluza:2016fwh}. 
%{\bf 
%by searching for a stationary phase contours $\mcC$  as solutions of properly defined differential equations. 
%}
%
%Here we focus on one-dimensional cases. In principle, the proposed here procedure can be applied to multi-dimensional MB integrals, though additional technical problems appear which need separate studies.  

Let us write a general \MB{} integrand $F(z)$, transformed into exponential form:\footnote{For brevity, we suppress the dependence on $s$ 
and shall use $F(z)$ instead of $F(s,z)$.}  
 \beq\label{Is}
I(s)=\frac{1}{2\pi i}\int\limits_{\mcC_0}
%c_0-i\infty}^{c_0+i\infty}
dz\,F(z)=\frac{1}{2\pi i}\int\limits_{c_0-i\infty}^{c_0+i\infty}dz\,e^{-f(z)}.
\eeq
$\mcC_0$ is a contour
% parallel to the $y-$axis 
defined by $\mathrm{Re}\,z=c_0$ while $f(z)=-\ln F(z)$.
%
%Typically MB integrands factorize into kinematical part $(-s/m^2)^{-z}$, which depends on Mandelstam variable $s$,  and a product of gamma and polygamma functions. Both parts exhibit 
%numerical problems 
%highly-oscillatory behaviour 
%when one tries to integrate along $\mcC_0$ \cite{jgll2016}. 
%
%The core of the problem with integration over $\mcC_0$ is highly-oscillatory behaviour of the integrand $F(z)$. 
%For such a class of integrands, standard methods of numerical integration 
%fail
%are often not adequate.

\begin{figure}[h!]
   \centering
   % \vspace{0.4cm}
     \includegraphics[width=.6\textwidth]{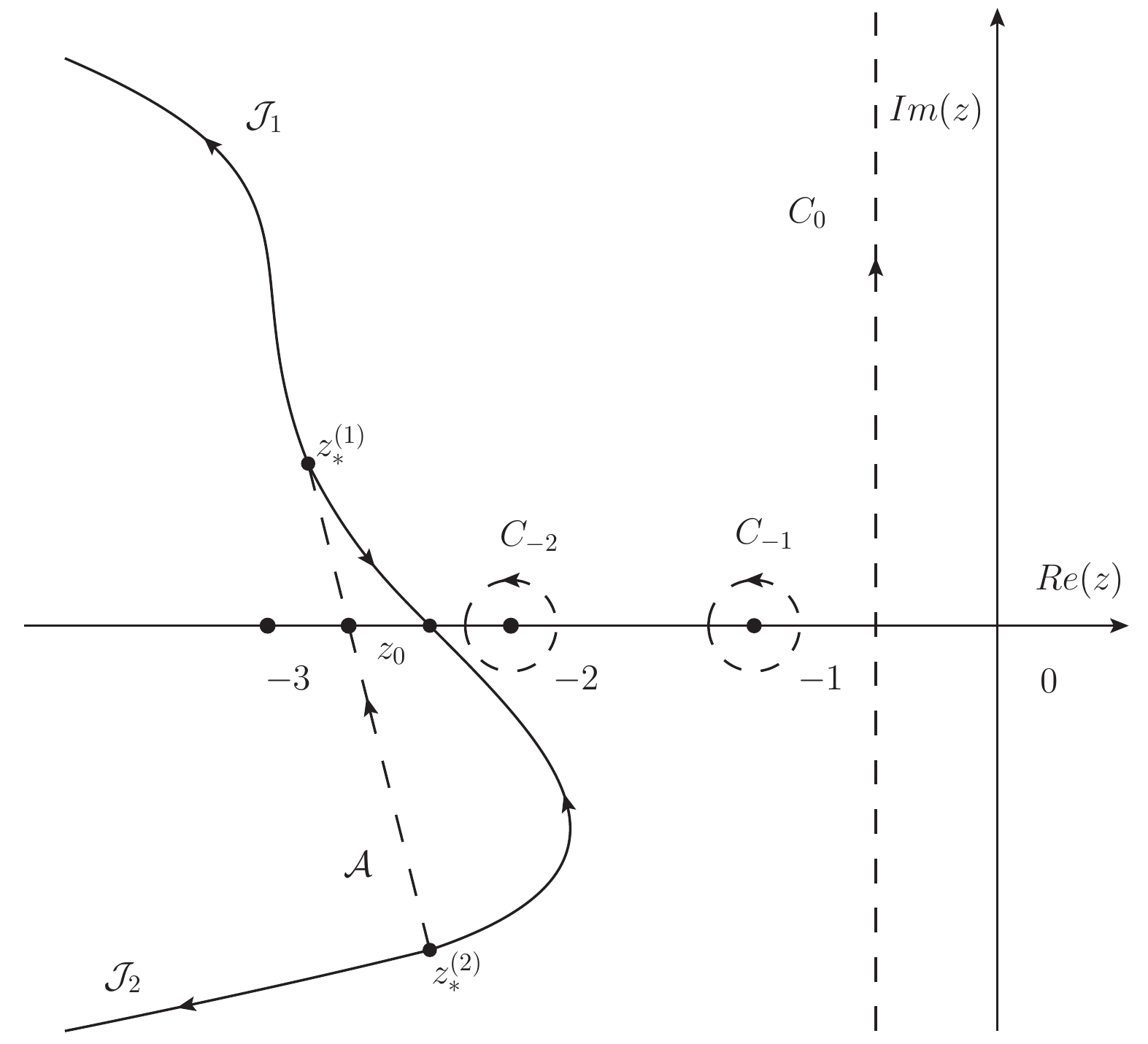}
\caption{
A deformation of the integration contour 
$\mcC_0$ defined by $\mathrm{Re}\,z=c_0$ to a contour $\mcC=\mcJ_1+\mcJ_2+\mcA$.
$\mcJ_{1,2}$ are two Lefschetz thimbles which start at saddle points $z_*^{(1,2)}$ and go towards infinity. 
The compact contour $\mcA$ (interval) connects the two saddle points $z_{*}^{(1)}$ and $z_*^{(2)}$. 
When there is an obstruction in deriving the parameterization of $\mcJ_{1,2}$ around some point, e.g. $z_0$, one can bypass that region 
using the contour $\mcA$. 
Note that here a deformation $\mcC_0\to \mcC$ requires taking into account integrals over two `small' contours, $\mcC_{-2}$ and $\mcC_{-1}$ around poles $z=-2$ and $z=-1$ which contribute to $\sum\res F$ in \eqref{lef1}. 
}\label{contours}
   \end{figure}

One of possible ways to 
%deal with 
get rid of 
 numerical problems with the \MB{} integrand $F(z)$ which is of  highly-oscillatory behaviour \cite{Dubovyk:2016ocz} is to integrate \eqref{Is} over a new contour 
$\mcC=
%\sum_{k=\pm}\mcJ_k
\mcJ_1+\mcJ_2+\mcA$. 
A typical example is sketched in Fig.~\ref{contours} where $\mcC$ is a sum of three contours $\mcJ_1$, $\mcJ_2$ and $\mcA$  along which 
the behaviour of $f$ is under control. 

Taking $f=\mathrm{Re}\,f+i\,\mathrm{Im}\,f$, we deform the original integration contour $\mathcal{C}_0$ to a Lefschetz thimble $\mathcal{J}(z_*)$, 
  \begin{equation}
 \int_{\mathcal{C}_0} dz\,e^{-f}=\overbrace{e^{-i\,\mathrm{Im}\,f|_{\mathcal{J}(z_*)}}}^{\rm Overall\; factor}
 \overbrace{\int\limits_{\mathcal{J}(z_*)}dz\,e^{-\mathrm{Re}\,f}}^{\rm Damping\; factor}
 +\overbrace{2\pi i\sum\limits_{\mathcal{C}_0\to\mathcal{J}(z_*)}\mathrm{Res}\,e^{-f}}^{\rm Remnants}.
 \label{lef1}
  \end{equation}
% }
 \\
The analytical formula describing $\mcJ_k$ can be found only in the simplest cases by explicit solving the equation 
$\mathrm{Im}\,f=\mathrm{const.}$ Instead, we use the fact that the  function $\mathrm{Re}\,f$ defines a Morse
% or Morse-Smale/gradient 
flow \cite{Nicolaescu:2011,Arnold:2012}.  
Such a flow is realized by a parameterization $t\mapsto z(t)$ of $\mcJ_k(z_*)$ in a form of Lefschetz thimbles 
\cite{Pham:1983,Witten:2010cx,Witten:2010zr,Harlow:2011ny,Kanazawa:2014qma,Tanizaki:2014xba}.
The {{Lefschetz thimble $\mathcal{J}(z_*)$ is defined as a union of curves $t\rightarrow z(t)=(z_1(t),\ldots,z_i(t),\ldots,z_n(t))\in 
\mathbb{C}^n$ which satisfy  the following differential equation  \cite{Witten:2010cx,Kanazawa:2014qma}}}:
 \begin{equation}
 \frac{d  z_i(t)}{dt}\,\,\,\,\,\,\,\,=\,\,\,\,\,\,\,\,-\left(\frac{\partial 
  f(z)}{\partial z^i}\right)^*,\quad z(+\infty)=z_*.
  \label{lef2}
\end{equation}
Here {{$z_*$ is a saddle point of a  meromorphic function $f$.
 {The crucial observation is that for  $\mathcal{J}(z_*)$ we can take  $\mathrm{Im}\,f$ = const, leading to the overall factor in 
(\ref{lef1}).
Note that  $\mathrm{Im}\,f$ generates  {a Hamiltonian flow} on $\mathbb{R}^{2n}$, e.g. for $n=1$:
\begin{equation}
\frac{d x(t)}{dt}=\frac{\partial \mathrm{Im}\,f}{\partial y},\quad \frac{d y(t)}{dt}=-\frac{\partial \mathrm{Im}\,f}{\partial x}.
\end{equation}
 
The  $\mathrm{Re}\,f$ is monotonically decreasing when $t\rightarrow+\infty$ and goes to $+\infty$ when $t\to-\infty$, leading to the 
damping factor in (\ref{lef1}).   
   
Remnants in (\ref{lef1}) are, according to Cauchy's theorem,  residues over poles when the integration contour is deformed from $\mcC_0$ to 
$\mcC$ and encircles extra  poles of $F=e^{-f}$, as shown in Fig.~\ref{contours}. 
In this way, (\ref{lef2}) gives a possibility to solve for $z(t)$ such that the integral (\ref{lef1}) is under control.
 As $\mcJ_{1,2}$ we choose such stationary phase contours 
 %$\mcJ_{1,2}$ are two contours 
 which start at saddle points $z_*^{(1,2)}$ and go towards infinity without hitting other poles. Both contours are chosen such that  
$\mathrm{Im}\,f$ is constant along them and function $\mathrm{Re}\,f$ is strictly increasing when one moves away from $z_*^{(1,2)}$. 
 Varieties defined in such a way are called   steepest descent contours
 \cite{Bender:1999,Wong:2001}. Usage of $\mcJ_{1,2}$ allows to control the behavior of $f(z)$ when $z\to\infty$. Because $\mathrm{Re}\,f$ 
is stricly increasing, the integrand $e^{-f}$ decreases rapidly at the ends of $\mcJ_{1,2}$. That transform the integral \eqref{Is} into a 
form which is more suitable for numerical treatment. 
 %makes the integral \eqref{Is} well-defined and fast convergent {\bf [not for all $f$; only for some appropriate $f$'s]}.

With respect to various methods known in the literature \cite{Anastasiou:2005cb,Czakon:2005rk,Dubovyk:2016ocz,Freitas:2010nx} which 
shift/rotate contours or use approximate forms  thereof, the \MB{} method which relies on the  differential equation (\ref{lef2}),  in 
short the \tt MBDE} method, relies on deriving the numerical parameterization $z(t)$ of $\mcJ_k$ 
%with the help 
as a solution of the differential equation \eqref{lef2} and then, again numerically, integrating the function  $e^{-\mathrm{Re}\,f}$ along 
the contour  $\mcC$ composed of Lefschetz thimbles $\mcJ_k$ (and the compact contour $\mcA$ if necessary). 
%Exact realization of a situation as in Fig.~\ref{contours} can be found in Fig.~... in \cite{Gluza:2016fwh}.
%
The purely numerical approach  {\tt MBDE} is complementary to the Pad\'e approximation presented in \cite{Gluza:2016fwh}.  

Let us shortly discuss numerical features of the {\tt MBDE} metod and display results of some  performance tests. We stress that the tests 
are preliminary, implemented directly in {\tt Mathematica}, in graphical mode, on an i7 $2.9$ GHz CPU. 
% -- see Tab.~\ref{tabF1}.
%Let us point out the main features of the method:
%
%\begin{itemize}
%
%\item 
Both kinematical regions $s<0$ and $s>0$ are treated in the same way in {\tt MBDE}, although $s>0$ seems to be more CPU time consuming.  For 
a final accuracy of the order of $10^{-6}$ the method  is as fast as $\texttt{MB}$ \cite{Czakon:2005rk} and $\texttt{MBnumerics}$ 
\cite{jgll2016}, while for an accuracy of $10^{-11}$ and higher, \texttt{MBDE} turns out to be more than 10 times slower than other two 
packages; see Tab.~\ref{tabF1}. To get a precision of the order of $10^{-16}$ some kind of optimization of the method is needed. 
Presumably, it can be made much faster by implementing in e.g. Fortran or C/C++ or by applying a dedicated method of solving 
differential equations. Parallelization or dividing integration regions into smaller parts to achieve larger precision are also possible 
options.

\begin{table}
\begin{center}
\begin{tabular}{ | c | c | c  | c | c | c |}
\hline
$s$ & $I_{\texttt{MBDE}}$ & $-\log_{10}\delta_{\mathrm{an}}$ & $T_{\texttt{MBDE}}[\textrm{s}]$ & $T_{\texttt{MB}}[\textrm{s}]$ & $T_{\texttt{MBnum}}[\textrm{s}]$\\
%\hline
%$-20$ & & & & & & 1.58 & 1.60\\ 
%\hline
%$-2$ & & & & & & 1.47 & 1.67\\ 
%\hline
%$-1$ & & & & & & 1.44 & 1.69\\ 
\hline
$-1/20$ & $4.96\times10^{-2}$ & $6/9/11$ & 1.31/2.48/16.94 & 1.43 & 1.16\\ 
\hline
$1+i0^{\pm}$  & -1.21 & 6/11 & 15.05/53.19 & -- & 1.28\\ 
\hline
$5$  & $4.30+14.05i$ & 10 & 13.5 & -- & 1.57\\
\hline
\end{tabular}
\end{center}
\caption{Performance tests of {\tt MBDE} for the integrand $F_1(z)=\left(-s\right)^{-z}\Gamma^3(-z)\Gamma(z+1)/\Gamma(-2z)$. 
The relative error  $\delta_{\mathrm{an}}$ is defined as $\delta_{\mathrm{an}}=|(I_{\mathrm{an}}-I_{\texttt{MBDE}})/I_{\texttt{an}}|$. 
$I_{\mathrm{an}}$ is the analytical value of the integral $I_1(s)$,  $I_{\texttt{MBDE}}$ is the numerical value of this integral evaluated 
with the {\tt MBDE} method. Finally, $T_{\texttt{MBDE},\,\texttt{MB},\,\texttt{MBnum}}$ display runtimes (in seconds) needed to numerically 
evaluate an integral using \texttt{MBDE}, \texttt{MB.m} and \texttt{MBnumerics.m} (with default settings), respectively. 
%following options: $\texttt{PrecisionGoal}\to4$, $\texttt{AccuracyGoal}\to12$, $\texttt{MaxPoints}\to10^6$, $\texttt{MaxRecursion}\to10^3$.
}\label{tabF1}
\end{table}
 
\section{Summary and Outlook}

In the last few years there is substantial progress in the direct calculation of multiloop integrals (Feynman diagrams)  in the physical, 
Minkowski regime using both SD and \MB{} methods.
 Both methods are complementary in several respects.
 In the \MB{} case, the most advanced is the top-bottom approach implemented in the \mbn{} package where multidimensional \MB{} integrals can be solved in physical kinematics with high accuracy for \MB{} integrals of dimension eight and below. 
 %the   In 2016, Mellin-Barnes methods developed for the same goals\\ $\longrightarrow$ MBnumerics
 Potential applications of the discussed numerical methods are complete 2-loop electroweak {pseudoobservables} 
needed for future linear colliders -- multi-massive 2-loop vertices -- and also non-resonant two-loop box diagrams -- complete cross 
sections, including LHC problems \cite{Freitas:2010nx})}.  
 
 { Using {\it {numerical}} methods, we are approaching {\it{automation}} in  calculation of Feynman {integrals} {\it{beyond}} the NLO level {\it{directly}} in {\it{physical}} kinematics. } Perspectives are robust, concerning both high and low energy physics. 

\section*{Acknowledgements}
\textit{I.D.}\ is supported by a research grant of Deutscher Aka\-de\-mi\-scher Austauschdienst (DAAD) and by Deutsches 
Elektronensychrotron{} DESY; 
%\\
\textit{J.G.}\ and \textit{T.J.}\ are supported by the Polish National Science Centre (NCN) under the Grant Agreement No. DEC-2013/11/B/ST2/04023;
%\\
{\textit{T.R.}\ is supported in part by an
Alexander von Humboldt Polish Honorary Research Fellowship.
The work of \textit{J.U.}\ is supported by Graduiertenkolleg 1504 ``Masse, Spektrum, Symmetrie`` of Deutsche For\-schungsgemeinschaft (DFG).}

%%%%%%%%%%%%%%%%%%%%%%%%%%%%%%%%%%%%%%%%%%%%%%%%%%%%%%
%\providecommand{\href}[2]{#2}
% \bibliographystyle{utphys_spires} % for Proceedings volume
%\bibliographystyle{elsarticle-num} % for hep-ph and so on, long version
%\bibliography{2loops_ambre2015}
\end{document}